# Synergetic Effect of Wall-Slip and Compressibility During Startup Flow of Complex Fluids

Aniruddha Sanyal, Sachin Balasaheb Shinde, Lalit Kumar*

*Department of Energy Science and Engineering, Indian Institute of Technology Bombay, Mumbai 400076, Maharashtra, India*

*Corresponding Author: lalit.kumar@ese.iitb.ac.in

ORCID ID: orcid.org/0000-0002-1946-8231

The present letter explains the synergetic effect of wall-slip, compressibility, and thixotropy in a pressurized flow startup operation of various structured fluids. Opposite to the intuition, experimental and numerical simulations suggest that the wall-slip (adhesive failure) is facilitating gel degradation (cohesive failure), revealing a new flow-startup mechanism. The thixotropic rheological model includes structural degradation kinetics at the bulk. Whereas, a static slip-based model addresses the near-wall phenomenon. The near-wall transient variations in axial velocity or strain evolution, and the initial pressure propagation mechanism along the axis of the circular pipe explain the essence of the aforementioned synergy.

Shear-induced forces during flow startup operation in a pipeline carrying complex fluids cause structural disintegration through compression, creep, shear-stress-localization, shear-banding, and hammering. Wall-slip occasionally instigates flow startup when the shearing strength is low [1-3]. Transportation of complex fluids like waxy crude oil gel, polymeric melts, paints, toothpaste, sewage waste, foodstuffs, and several suspensions or emulsions show such wall-slip effects during flow startup [4-6].

Some high molecular-weight organic compounds characteristically disobey hydrodynamic no-slip at the fluid-wall interface (FWI) beyond a certain stress $\tau_c$ (often termed as "sliding yield stress" [7] or "critical stress" for wall-slip [8]) during flow initiation. The wall-slip in a pipe is a shear-dependent phenomenon wherein velocity discontinuities at the wall accounts for highly sheared thin region adjacent to the wall having very low viscosity compared to the bulk [9, 10]. Wall-slip reduces the yield stress requirement at the FWI (as seen for colloidal silica gels [11]) without changing the rheological properties of the fluid [11, 12]. The rheology at the FWI is governed by its surface properties during flow startup operation [13]. In the case of polymer melts, Brochard & De Gennes [14] interpreted the interface as a region grafted with few chains identical to polymer melt flowing in bulk. Above $\tau_c$, the grafted chains undergo a coil stretch leading to disentanglement and subsequent slippage. Wall-slip may also happen when gaseous films are at the FWI or where water flows through hydrophobic capillaries [15]. Consequently, the velocity discontinuity at the wall is a common feature of the wall-slip phenomenon in all these scenarios (Mooney [16] was the first person to report this).

The wall-slip causes flow instability, resulting in non-linear dynamics (quasi-periodic and chaotic flow) at the FWI. According to Graham & Coworkers [5, 8], the stress history at the wall-boundary influences this instability, and one should incorporate it in the slip-based rheological model. Spikes et al. [10] broadened this shear stress-based criteria using critical wall-shear stress at which the slip begins, thereafter, confined within a constant slip-length. The overall deformation in the fluid's structure at the wall is quantitatively explained by the apparent shear rate $\dot{\gamma}_{app}$ which is the combined effect of the nominal shear rate due to bulk flow $\dot{\gamma}_n$ and the slip at the wall ($u_s/b$). At low shear rates, $\dot{\gamma}_{app}$ is only due to the surface effects [6, 8, 11, 17-19]. The deformation due to slip is shown to be a power-law function of the shear stress at the wall (discussed more in detail in the Methods section).

In startup flow, the complex fluid initially ruptures due to an adhesive failure at the wall resulting from the shearing confinement. At a later stage, continuous shear deformation results in cohesive failure (or disengagement) [20-21]. One can expect the wall-slip, through adhesive failure, initiates complex gel movement from inlet to outlet of a pipe at the smallest time scale of flow, i.e., during initial pressure propagation (IPP). The initial gel rupture mechanism can have a lasting effect on the flow startup operation, as it dictates the pressure gradient in the subsequent section of the pipeline [22-25]. For example, the wall-slip phenomenon prevails during the flow assurance of waxy crude oil pipelines at subsea conditions. Literature indicates that waxy crude oil, similar to other complex fluids, can exhibit different phenomenological or indirect-microstructure-based complex rheology [21-23, 26-31]. The investigations on flow startup using weakly compressible waxy crude oil gel can create a benchmark analysis for operations involving a larger group of complex fluids.

Flow startup operation is theoretically best understood through initial compressive pressure wave propagation for complex fluids and subsequent shear-layer development, leading to destructing of the complex fluids structure [22, 24, 32-35]. During IPP the pressure gradient is generally high at the compressive pressure front (CPF) in most parts of the pipeline. The high local pressure gradient at the front may cause an adhesive gel failure, resulting in slip flow. In theory, the wall-slip effects during adhesive breakage remains unexplained for transient compressive pressure wave movement. The slip can result in the un-attenuated propagation of pressure signals along the pipeline axis. It intuitively indicates that a high pressure gradient may not result in shear deformation and subsequent de-structuring of the fluid. However, the intuition of low overall structural degradation compared to the no-slip scenario is far from true for most complex fluids. The analysis involving IPP phenomenon must address the contribution of wall-slip in rheological formulations for correct assessment of the flow behavior.

**Results**

The mechanism for elasto-hydrodynamic slip at the interface of a soft fluid or glassy material and wall-surface has been comprehensively studied in the literature [7, 36]. However, the wall-slip effects during compressive pressure propagation and gel degradation for flow startup remains unknown. Hence, we carefully examine the combined role of wall-slip and compressibility in various complex gel degradation processes during startup flow. Initially, experiments are performed to decipher the wall-slip effects on the gel degradation mechanism at the bulk. A model oil with 10% wax concentration is cooled from 45°C to 4°C with a cooling rate of 1 °C/min. Following a 10 min hold, the sample is subjected to a constant stress of 100 Pa until the material breaks (or 1 hour whichever is earlier). The results are compared for the cases with smooth and rough inner surfaces (the exact parameters for smoothness and roughness are discussed in "Methods" section). Counter-intuitively, one may see that the smooth surfaces show increasing gel deformation quantified through strain parameter compared to negligible deformation for the rough surfaces. Our preliminary numerical analysis, as discussed hereafter, gratifies the experimental outcome (Figure 1b).

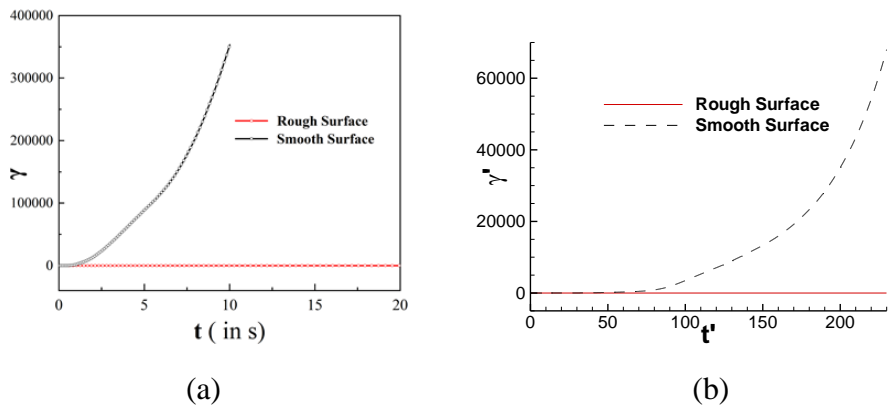

**Figure 1. Comparison of strain evolution showing flow and no-flow scenario for smooth surfaces (signifying wall-slip) and rough surfaces (signifying no-slip) using (a) experimental (for parallel plate configuration in Anton Paar MCR 301 rheometer) and (b) numerical simulations (at a location near the inner wall of a pipeline).**

**Model Development:** As a schematic, we consider a horizontally aligned cylindrical pipeline clogged homogeneously with elasto-viscoplastic or shear-thinning-based thixotropic fluids (e.g., waxy crude oil gel). An isothermal startup operation of the pipeline is initiated by applying pressure $P$ at the inlet using a Newtonian fluid having a property equivalent to that of the complex fluids at completely destructed state. The compressibility of the gel $\kappa_\Theta$ vary between $10^{-10}$ Pa$^{-1}$ to $10^{-7}$ Pa$^{-1}$, signifying nearly-incompressible and moderately-weak compressibility limits [22, 24]. The present numerical study assumes an axisymmetric domain $\Omega$ in the range $[0, L] \times [0, R]$ in polar coordinate system $(r, \theta, z)$.

The following constitutive functional form is used to represent tangential slip velocity at the wall:

$$u_s = \emptyset(\tau) = \begin{cases} 0, & |\tau| < \tau_c \\ B(|\tau| - \tau_c)^m, & |\tau| \geq \tau_c \end{cases} \dots\dots\dots\dots\dots (1),$$

where $m$ is a power-law parameter governing slip. The variable $B$ depends on kinetic parameters, and for isothermal study it is a constant [37, 38]. When shear-thinning-based slip is possible, $\tau_c$ becomes 0 [32]. However, for the yielding fluids, we have considered partial slip where $\tau_c$ becomes some fraction of the yield stress $\tau_y$ (e.g., $\tau_c = 2/3\,\tau_y$, estimated from the experimental probes for various complex fluids, as shown in the "Methods" section). The best-suited numerical values for $B$ and $m$ are finalized after analyzing various fluids through rheometric experiments (details are provided in "Methods" section).

The problem is defined through the conservation principles for mass and momentum [25] along with strain evolution equations to assert the coupling of structural degradation-based kinetics with the constitutive model for extra stress tensor $\ddot{\boldsymbol{\tau}}$. The extra stress tensor $\ddot{\boldsymbol{\tau}}$ is represented in terms of viscous and elastic components of stress as follows:

$$\ddot{\boldsymbol{\tau}} = \mu((\nabla \boldsymbol{U}) + (\nabla \boldsymbol{U})^T) - \frac{2}{3}\mu \nabla \boldsymbol{U}\, \ddot{\boldsymbol{I}} + G\ddot{\boldsymbol{\gamma}} \ldots\ldots\ldots(2),$$

with $G$, $\ddot{\boldsymbol{\gamma}}$, and $\mu$ being the effective elastic modulus, nominal strain tensor and effective gel viscosity, respectively. Finally, the evolution of $\ddot{\boldsymbol{\gamma}}$ for infinitely high relaxation time is represented in frame-invariant upper convection derivative form as follows [24, 39]:

$$\breve{\ddot{\boldsymbol{\gamma}}} = \frac{\partial \ddot{\boldsymbol{\gamma}}}{\partial t} + \boldsymbol{U} \cdot (\nabla \ddot{\boldsymbol{\gamma}}) - (\nabla \boldsymbol{U})^T \cdot \ddot{\boldsymbol{\gamma}} - \ddot{\boldsymbol{\gamma}} \cdot (\nabla \boldsymbol{U}) = ((\nabla \boldsymbol{U}) + (\nabla \boldsymbol{U})^T) \ldots\ldots\ldots (3),$$

where $\breve{\ddot{\boldsymbol{\gamma}}}$ is the upper convection derivative. The magnitude of strain tensor $\ddot{\boldsymbol{\gamma}}$ can be determined after solving for each component of the strain tensor ($\gamma_{rr}$, $\gamma_{\theta\theta}$, $\gamma_{zz}$, $\gamma_{rz}$) in Eq. (3) using the following form:

$$\gamma = \|\ddot{\boldsymbol{\gamma}}\| = \sqrt{\gamma_{rz}^2 + \frac{1}{2}(\gamma_{rr}^2 + \gamma_{\theta\theta}^2 + \gamma_{zz}^2)} \ldots\ldots\ldots (4).$$

The structural degradation-based constitutive model for stress is written as:

$$\tau = \mu_\infty \left(1 + \frac{\mu_r}{(1+2k\gamma)^{1/2}}\right)\dot{\gamma} + \frac{G_0}{(1+2k\gamma)^{3/2}}\gamma \ldots\ldots\ldots (5),$$

where $\mu_r$ is the viscosity ratio ($= \mu_{g0}/\mu_\infty$). Eq. (5) follows the flow curve at varying values of constant strain rate $\dot{\gamma} < 10$ s$^{-1}$ in the experimental results of Zhao et al. [40, 41]. In the creep flow regime, the major contribution for stress comes from the elastic response. The value of strain at maximum stress in such condition is considered as "yield strain", which depends on the degradation rate constant (i.e. $\gamma_{max} = 1/k$). The static yield stress $\tau_y$ can be determined at $\gamma = \gamma_{max}$ in Eq. (5), which turns up to be: $\tau_y = G_0/3^{3/2}$. The present constitutive model consistently replicates the patterns from the literature [40-42] e.g., overshoot in a shear-rate-controlled stress-strain flow curve, shear hysteresis and shear banding. For shear-thinning fluids, the assumption of a high shear rate makes the second term on the right-hand side of Eq. (5) negligible. However, for a material showing Kelvin-Voigt type response to loading, the viscosity and elastic-modulus are time and strain-independent. The present letter purposefully avoids mechanical analogs which capture yielding behavior using extreme limiting parameters (e.g. infinite viscosity [43]). One may note that the structural degradation kinetics considers that the timescale for gel-breakdown or IPP during startup is much smaller than the gel-buildup timescale (also realistic for waxy crude oil gel [44, 45]). Eq. (5) is derived following the mathematical sequences stated by Cheng & Evans [46]. The equations are scaled in terms of the aspect-ratio of the pipeline ($1/\varepsilon$), the timescale capturing compressive pressure movement $t'$ and the ratio between the actual pipe length to the critical length $\alpha$ signifying the pressure force to wall-shear related force balance (discussed in detail in "Methods" section). In this problem, FVM-based methodology is applied in a numerically adequate staggered-type orthogonal grid setup to solve for primary variables like $U$, $p$ and $\ddot{\gamma}$ using the point-by-point iterative method (verification of the algorithm is provided in point 1 of the supplementary material). The symbol "'" throughout indicates the dimensionless forms.

**Analysis:** The gel deformation and subsequent gel degradation mechanisms during wall-slip can be best understood by determining the physical phenomena occurring near the wall at various time instants. The time evolution of local axial velocity $w'$ near the wall ($r' = 0.975$) at $\alpha z' = 0.1$ (Figure 2a) indicates the scenario in which the pressurized fluid has just entered the pipe from the inlet. At compressibility number $\delta = 4 \times 10^{-4}$ ($= P\kappa_\theta$), the acoustic speed $w_f$ ($\approx 1/\sqrt{\rho_0 \kappa_\theta}$) for initial "compressive" pressure propagation is 333 m/s. Based on this, the CPF will travel a 7.5 m long pipeline in $t' = 2.252$. The minimum time for the pressure signal to reach $\alpha z' = 0.1$ is $t' = 0.15$. Furthermore, the gel starts degrading substantially at $t' \approx 9.16$ causing a sharp rise in velocity. At this juncture, the applied pressure overcomes the net viscoelastic force across the wall (this concept was explicitly explained for a no-flow startup [31]). One can subsequently note the earliest hint of flow from the outlet at $t' \approx 9.16$ due to inertial puncture [23] (Figure 4 in the supplementary material). However, the density barrier at the outlet of the pipe causes reflection of pressure waves through the impedance phenomenon [47]. Additionally, an increase in $w'$ is noted accompanied by continuous reduction in local elastic forces for $t' > 9.16$. This understanding is consistent with the strain evolution pattern at $\alpha z' = 0.1$ for $t' > 9.16$ (Figure 3a). The magnitude of local strain increases sharply at $t' \approx 15$ (Figure 3a) along with a sharp rise in $w'$, as observed in Figure 2a. The local near-wall elastic forces at $\alpha z' = 0.1$, and net elastic forces along the entire axial stretch of the wall decline to insignificance at $t' \approx 50$. Once the gel is uniformly compressed, one may see the inception of flow with a constant axial pressure gradient indicating uniform resistance to flow at all cross-sections. Thereafter, $w'$ monotonically increases with time to a steady-state at $t' > 300$. As per the established sequence for gel degradation where the accumulated strain diffuses radially inward with time [24], the sheared region spreads from the near-wall region to the interior bulk. The initial transient feature of $w'$ is caused by the viscous decaying of shear layers along with multiple flow reflections from the outlet to the inlet due to impedance [47]. Other than the fact that a sizeable portion in the upstream is already compressed and partially deformed, the flow characteristics (in terms of $w'$) along the wall at $\alpha z' = 0.75$ (Figure 2b) remains similar to that at $\alpha z' = 0.1$. The initial CPF reaches $\alpha z' = 0.75$ at $t' \approx 1.126$. The inlet-based influence on flow rearrangement at $\alpha z' = 0.1$ diminishes at $\alpha z' = 0.75$. Hence, $w'$ for wall-slip at $\alpha z' = 0.75$ remains higher during initial time (also verifiable from the time evolution of inlet and outlet flowrates (Figure 4 in the supplementary material)).

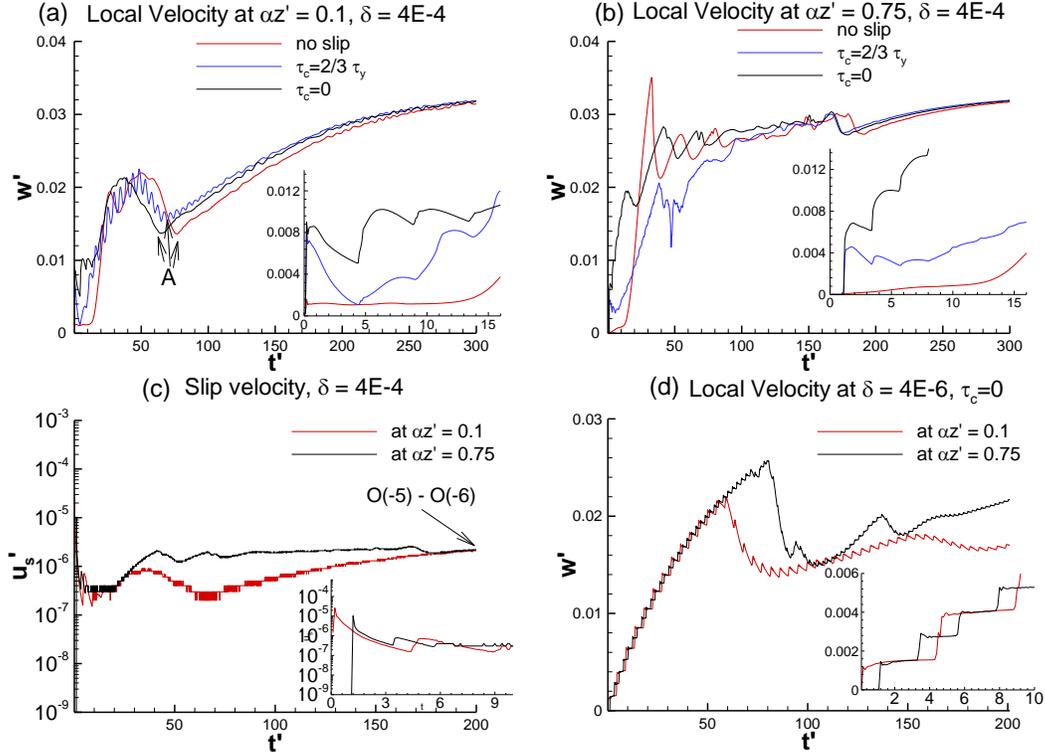

**Figure 2.** Comparison for time-dependent variations in local axial velocity $w'$ near the fluid-wall interface at $\mu_{g0} = 100$ Pa s, $\alpha = 1.5$, $P = 40$ kPa, $k = 100$, $\delta = 4 \times 10^{-4}$ at (a) $\alpha z' = 0.1$, and (b) $\alpha z' = 0.75$; with the cases for no-slip, conditional slip ($\tau_c = \frac{2}{3}\tau_y$) and shear-thinning slip ($\tau_c = 0$). (c) Variations in slip velocity with time at $\tau_c = 0$ for different axial locations. (d) Comparison for $w'$ at same gel condition with $\delta = 4 \times 10^{-6}$, at $\tau_c = 0$.

For wall-slip scenario, an altered compressional deformation and wall-stress causes increase in $w'$ with respect to no-slip. Here, the local flow commences when the CPF reaches $\alpha z' = 0.1$ and $\alpha z' = 0.75$, irrespective of wall conditions. The initial increment in $w'$ for wall-slip at $\alpha z' = 0.1$ is greater than that at $\alpha z' = 0.75$, indicating prominence of initial inertial compression near the inlet. Interestingly, the sudden rise in $w'$ comprises the cumulative effect of slip and no-slip velocities at the nearby wall, which initially support enhanced compressional deformation at CPF during IPP. Furthermore, if we refer to the inset of Figure 2a, we see a drop in $w'$ till $t' \approx 4.729$. From an apparent viewpoint, one may expect the initial CPF to propagate to the end of the pipeline by $t' \approx 2.252$ and reflect to the inlet by $t' \approx 4.504$. Once the information of "no-resistance" reaches to the inlet, the inlet flow increases as a part of recurrent process during pressure wave propagation. The additional flow resistances due to wall-yield critical stress at $\tau_c = {}^2\!/_3\, \tau_y$ leads to greater initial drop in $w'$ compared to $\tau_c = 0$. At very low compressibility ($\delta = 4 \times 10^{-6}$) with wall-slip, $w'$ increases depending upon axial position during IPP (Figure 2d). One may also note that the rate of increase in $w'$ in case of $\delta = 4 \times 10^{-6}$ is almost same irrespective of the axial location. Compressive deformation during IPP is negligible, and the flow is governed by the slip alone. In such case, a linear pressure profile develops before shear deformation becomes significant to counter applied pressure (as shown in Figure 5 of the supplementary material). This allows the gel to deform uniformly throughout the pipeline. With increasing time, as the overall gel starts deforming, the magnitude of successive hikes in the $w'$ subsides. However, once the entire upstream undergoes deformation, $w'$ at $\alpha z' = 0.1$ shows a hike, indicating accelerated flow. This is consistent with the plots for the time evolution of the strain (in Figure 3c). The time-evolution of strain in Figures 3(a, c) shows an instantaneous hike concomitant with the understanding from the local flow variations. One may note from Figure 3c that the nominal (or actual) strain $\gamma'$ for the no-slip scenario eventually becomes more than that for the wall-slip induced scenario. This suggests that while the wall-slip influences the initial strain, but once IPP is done with, the local strain due to overall deformation in a no-slip condition surpasses that of the wall-slip scenario. To verify strain characteristics, we have checked for time-evolution of $\gamma'$ at a nearby radially inward location (Figure 3b). One can clearly see that in all cases, strain away from the wall remains substantially lower than the nominal strain at $r' = 0.975$ for initial time instants. One expects that at the earliest instant when there is hardly any bulk deformation in the gel, including the locations near the wall at the inlet, the strain for no-slip condition should be more than the wall-slip induced scenario. This is precisely recovered at $\alpha z' = 0.1$ at $t' < 0.1$, when the pressure force induced at the inlet is limited to 20 kPa (Figure 3d). In addition, it may be shown that these counter-intuitive outcomes on wall-slip-effects reverses when the fluid is almost incompressible.

The non-periodic variations in $w'$ in Figures 2(a, b) symbolize the slip-stick mechanism at $\tau_c = {}^2\!/\!_3\, \tau_y$. This nonlinearity can be qualitatively explained by the wall-slip model [8]. The present rheological model accurately predicts decreasing strain rate during the initial stage of flow startup (a decrease in strain rate at earlier times during flow startup is due to "shear localization" [45]).

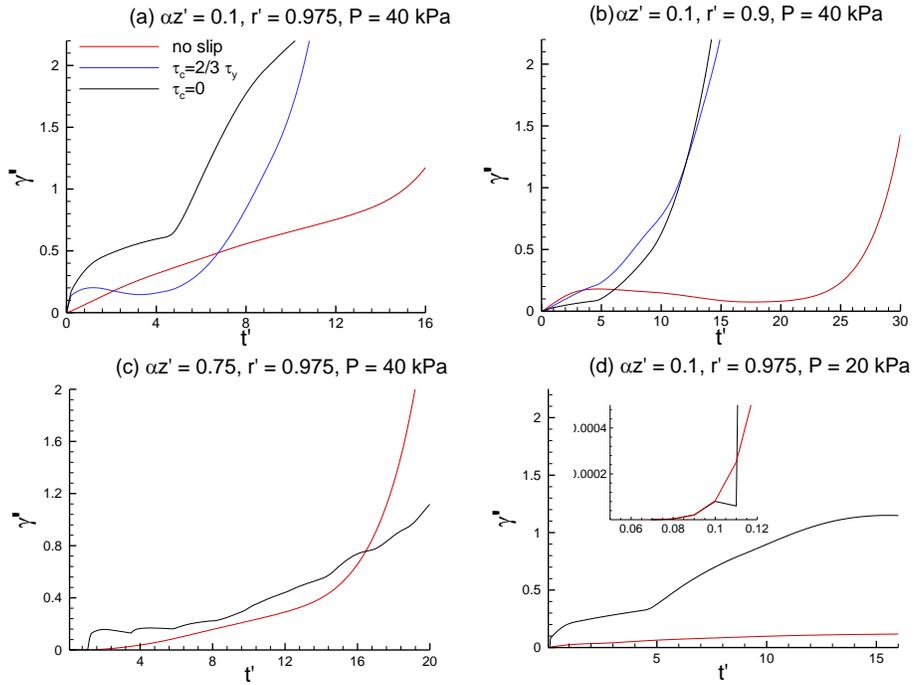

**Figure 3.** Comparison for the time evolution of strain $\gamma'$ between (a) no-slip, conditional-slip and shear-thinning slip at $r' = 0.975$, $\alpha z' = 0.1$ and $P = 40$ kPa, (b) at a radial location $r' = 0.9$, $\alpha z' = 0.1$, and $P = 40$ kPa, (c) no-slip and shear-thinning slip at $r' = 0.975$, $\alpha z' = 0.75$ and $P = 40$ kPa, and (d) no-slip and shear-thinning slip at $r' = 0.975$, $\alpha z' = 0.1$ and $P = 20$ kPa; while the other gel conditions remaining the same as in Figure 2.

Furthermore, we extend these understandings for other structured materials (like shear-thinning-fluids, Kelvin-Voigt (KV) type viscoelastic solid) to establish a benchmark overview to the complex flow physics associated with the synergetic effect of wall-slip and compressibility during flow startup. In the process, we also exhibit robustness of our model in the presence of wall-slip through predicting continuous flow.

For weakly compressible shear-thinning (ST) gel ($\delta = 4 \times 10^{-4}$) with no-slip, $w'$ (at $r' = 0.975$ and $\alpha z' = 0.1$) is initially higher than that of the thixotropic elasto-viscoplastic fluids (TEVP) (Figure 4a). Unlike TEVP, the applied pressure force for ST does not require to overcome elastic forces at the wall. For TEVP, $w'$ increases drastically after gel degradation (when $\gamma' > \gamma_{max}'$). The deformation in the present problem is directly associated with microstructural rearrangement of the gel's network guided through Eq. (5). For ST, the deformation in the bulk segment of the gel (except at the outset) during IPP is high compared to TEVP. Hence, at $t' > 21$, $w'$ for ST at the vicinity of the wall is affected by shearing actions between the adjacent shear layers along the bulk radial direction, which is evident from parabolic-type flow axial profiles. Whereas in TEVP, the flow occurs in the form of a plug due to strong shear bands in the bulk [24, 31]. During IPP with or without wall-slip, $\gamma'$ in the bulk portion of the gel remains less than the yield strain in TEVP. Besides, the wall-slip causes additional flow due to extra compression at the CPF by un-attenuated pressure force. However, the ST is only subjected to viscous shearing action, which causes continuous deformation in the bulk region of the gel. Unlike TEVP, the redistribution of energy for gel degradation is not just localized close to the wall but is widespread for ST. Consequently, the accumulation of shearing stress in the vicinity of the wall is lesser than that of the TEVP. Hence, a higher magnitude of $w'$ occurs for TEVP in Figures 4(b, c) (i.e. irrespective of the compressive resistances in the gel) during IPP.

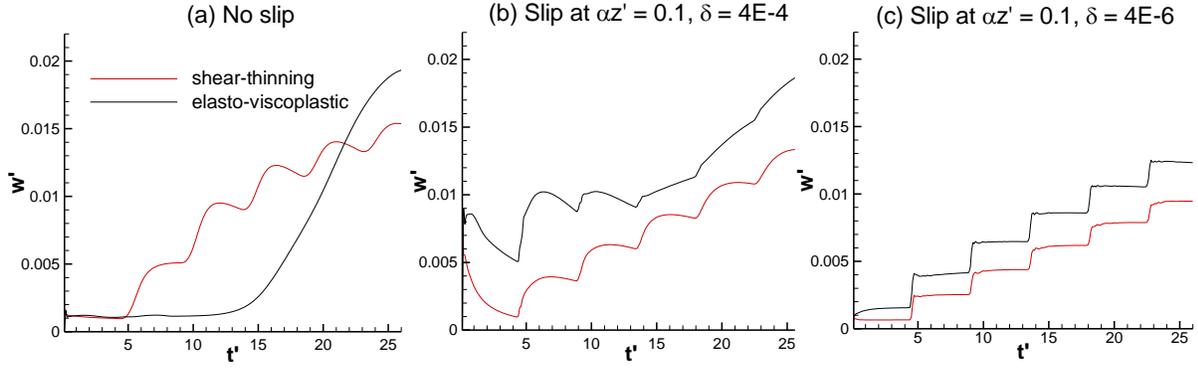

**Figure 4.** Comparison for time-variant local axial velocity for shear-thinning and elasto-viscoplastic fluids at $r' = 0.975, \alpha z' = 0.1, P = 40$ kPa for the case of (a) no-slip scenario at $\delta = 4 \times 10^{-4}$, (b) $\tau_c = 0$ at $\delta = 4 \times 10^{-4}$, and (c) $\tau_c = 0$ at $\delta = 4 \times 10^{-6}$.

For a weakly compressible KV-type material, the synergy between compressibility and wall-slip explains nature of pressure propagation and its continuous movement. The pressure profiles in Figure 5a show that the compressive resistances in the flow delay IPP. Despite the no-slip scenario, the CPF advances with initial inertial compression causing deformation near the wall at the inlet. This actuates small movement in the axial direction without sustainable flow. For wall-slip at $\tau_c = 0$, one may note higher $w'$ compared to no-slip cases after the subsidence of the initial flow transients triggered from inertia-based compression at the outset (Figure 5b). After a certain time, $w'$ attains a velocity having contributions from the bulk flow overcoming compressive resistances at the upstream (in addition to slip-velocity). In KV-type material with no-slip, the pressure will eventually balance by the elastic force and flow stops. However, unlike the no-slip scenario, the wall-slip allows continuous movement of the KV-type material. In the slip-flow, a higher pressure drop in the upstream portion of the gel setup during IPP is observed (Figure 5c). For the wall-slip scenario, overcoming the reduced wall-stress requirement is enough to push the KV-type material to the outlet in a stable plug-like format. Accordingly, a scenario occurs at some $\delta$ (within $4 \times 10^{-5} - 4 \times 10^{-6}$) where the KV-type material may flow to the outlet due to wall-slip, and the flow does not occur in a no-slip scenario. Thus, the wall-slip governs the flow of a viscoelastic-type solid material (like KV-type) at a later time during startup operation, opposite to what is seen for TEVP or shear-thinning fluids.

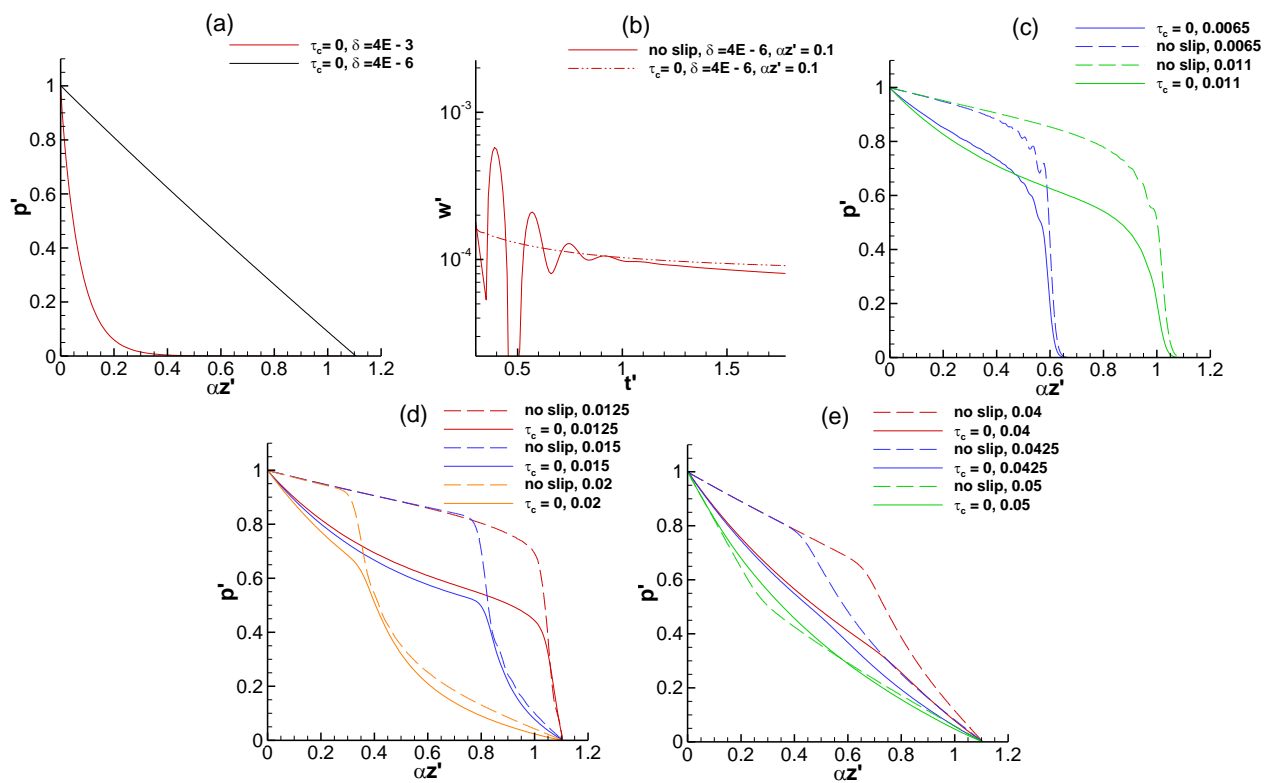

**Figure 5. (a) Effect of compressibility on flow startup during pressure propagation through a Kelvin-Voigt material at a later time $t' = 20$, $\alpha = 1.1$, and $\tau_c = 0$. (b) Transients in local axial velocity at $r' = 0.975$, $\alpha z' = 0.1$, $\alpha = 1.1$, and $\tau_c = 0$. Comparison for the time evolution of pressure propagation between no-slip and shear-thinning slip scenario ($\tau_c = 0$) for Kelvin-Voigt material (at $\delta = 4 \times 10^{-6}$) (c) $t' = 0.0065$ & $0.011$, (d) $t' = 0.0125, 0.015$ & $0.02$, and (e) $t' = 0.04, 0.0425$ & $0.05$.**

Finally, a comparison of pressure propagation at various time instants is shown in Figures 6(a-d) for cases involving slip and no-slip at the FWI for weakly compressible TEVP fluids. At an earlier time $t' = 0.1$, pressure builds up near the entrance due to inertial compression, as discussed earlier. The CPF diffuses further into the downstream at $t' = 1$ (Figure 6a). For the cases involving wall-slip, the viscous attenuation is less, and hence, the oscillations travel downstream. A steady decreasing pressure slope in upstream for a no-slip scenario indicates substantial bulk gel deformation. For wall-slip cases, this deformation is less, and the majority of the gel in the bulk region away from the wall remains intact at $t' = 1$. One may note that at $t' = 2.3$, the pressure propagates downstream at a higher speed for the wall-slip cases. In conclusion, the slip tends to dominate initial CPF movement during startup.

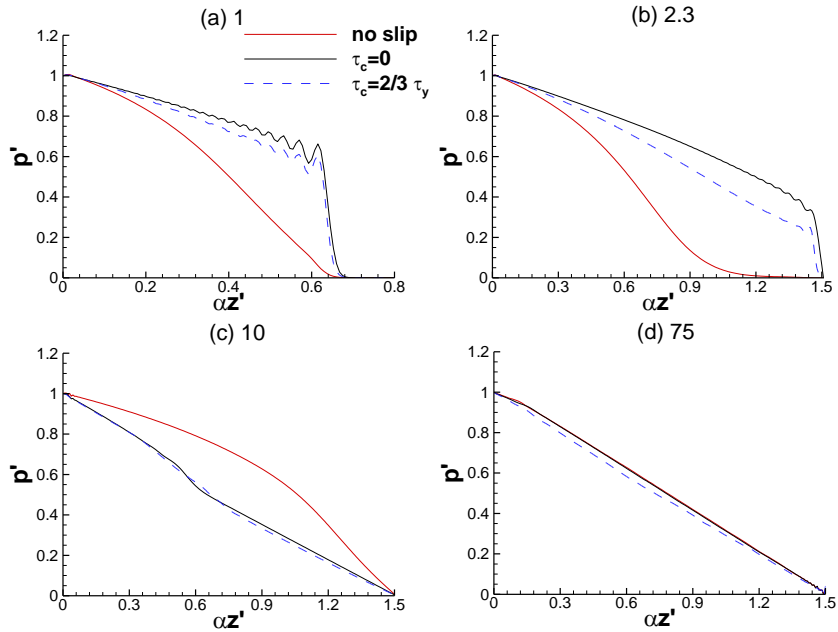

Figure 6. Comparison for the time evolution of axial pressure profile at $\mu_{g0} = 100$ Pa s, $\alpha = 1.5$, $P = 40$ kPa, $k = 100$, $\delta = 4 \times 10^{-4}$ with the cases for no-slip, shear-thinning slip ($\tau_c = 0$) and conditional slip scenario ($\tau_c = 2/3\ \tau_y$) at $t' =$ (a) 1, (b) 2.3, (c) 5, and (d) 75.

The pressure propagation mechanism at $\delta = 4 \times 10^{-6}$ sees multiple reflections of pressure waves from the outlet with slow gel deformation for a wall-slip scenario (the mechanism is explained in detail in point 3 of the supplementary material). For an energy efficient flow startup, pressure requirement estimations in the longer pipeline become intriguing. It is important to realize the importance of wall-slip in such scenarios. The present rheological model improves startup estimations for a longer pipeline ($\alpha = 4$) and low gel degradation rate constant ($k = 50$) (shown in point 4 of the supplementary material).

**Concluding Remarks:** This analysis creates a benchmark for any flow showing synergy between compressibility and wall-slip. The study can be extended to various structural degradation kinetics involving the effects of structural buildup (often realized for wormlike micelles). The analysis may be useful for demarcating the effect of shear banding in complex fluids where wall-slip inherently occurs [2]. To date, the concept of critical stress calls for bigger clarity on what yields and what might not yield. This letter hints at the importance for such clarity.

## METHODS

**Experimental determination of parameters of the slip-model (Eq. (1))**

**Materials**

Examinations are carried out based on complex fluids like model oil with 5-7% wax concentration (TEVP fluid), toothpaste (yield-stress fluid) and 1.5% Carbopol solution (Herschel-Bulkley fluid). The sample of model waxy oil is prepared by adding different macro-crystalline wax (Sasolwax 5054) concentrations ranging from 6 to 10 wt.% in Dodecane solvent. The sample of model oil was heated 10-20 °C above WAT to assure complete solubility of the wax in Dodecane. The Carbopol solution is prepared by adding 1.5 wt.% of Carbopol powder-940 in distilled water. This mixture is then rotated at 1100-1300 rpm for 30 min to ensure a homogeneous gel formation. In addition, commercially available toothpaste is used.

**Experimentation**

A series of rheological experiments are performed with the Anton Paar MCR 301 rheometer to investigate the wall-slip for complex fluids with yielding behavior. The fluid sample is kept on the fixed bottom plate of the rheometer. Two parallel-plate geometries (smooth and rough types) with 50 mm diameter are used for all rheological experiments (Figure 7). The surface roughness of the plates is measured using a Surface Profilometer: Alicona. The surface roughness of the smooth and rough plates is in the range of 1.5-2.2 $\mu m$ and 69.2-70.9 $\mu m$, respectively A constant gap of 1 mm prevails between parallel plates during measurements. A Peltier plate controller from the bottom plate maintains the temperature of the sample. For the case of waxy oil samples, the initial temperature is kept above WAT, and further, it is cooled to below the gelation temperature with a cooling rate of 1 °C/min. However, in the case of the toothpaste and Carbopol solution, an isothermal temperature of 25 °C is maintained throughout. After holding the sample for sufficient time, the yielding behavior of soft gelled fluid is investigated with the stress-ramp test. The stress ramp of 20 Pa/min for the case of toothpaste and waxy oil, and 6 Pa/min for Carbopol solution is applied during measurement. The shear stress corresponding to the sudden change in shear rate during the test is regarded as the yield stress of the sample. Additionally, the yielding behavior of the samples is investigated through the constant shear-rate method. A constant shear rate varying from 0.001 s$^{-1}$ to 10 s$^{-1}$ is applied to the gelled sample till complete degradation. Finally, the shear resistance offered by the material against deformation is recorded. The maximum shear stress in the constant shear-rate method is considered to be the yield stress of the material where it starts to flow.

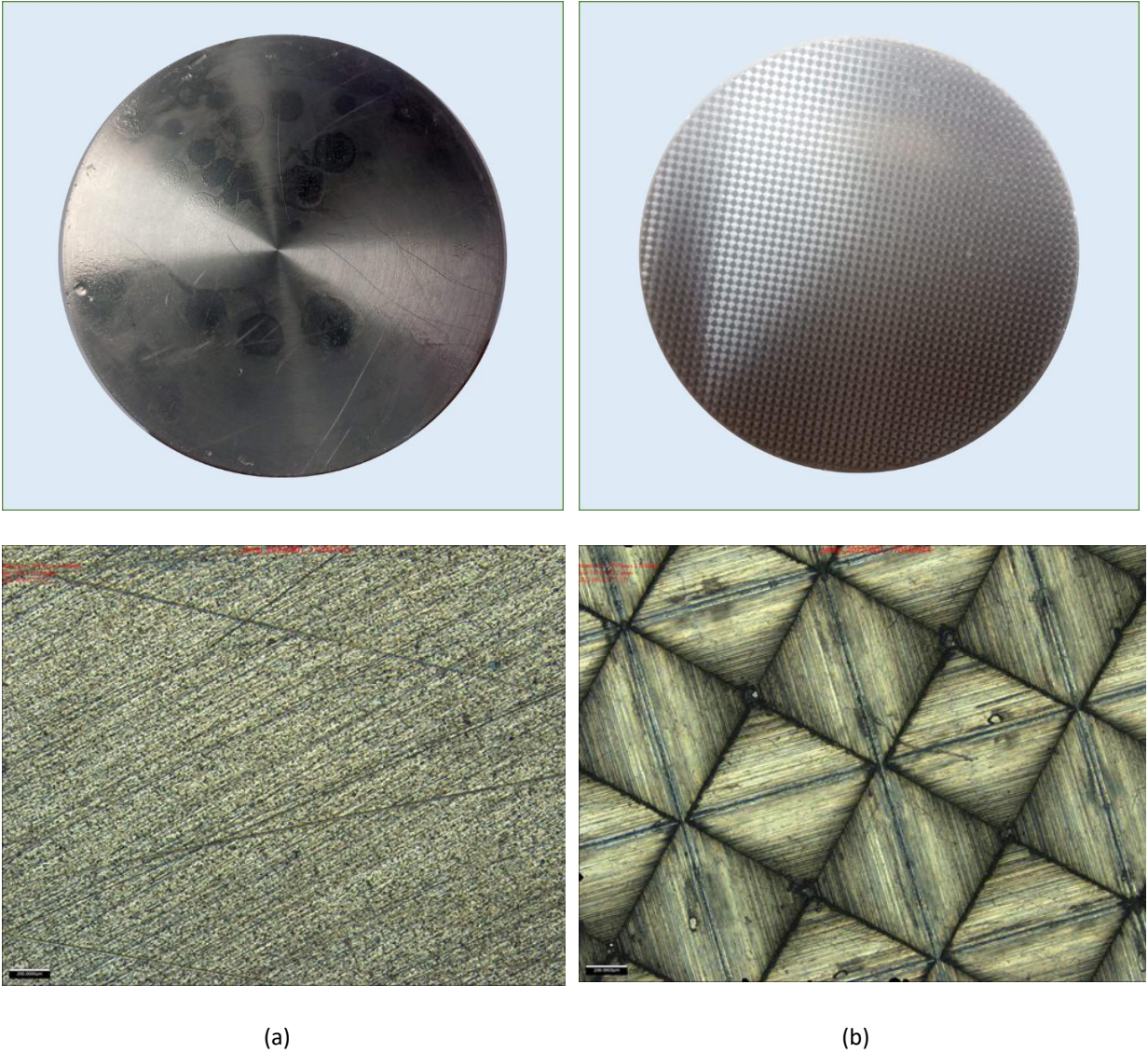

(a) (b)

**Figure 7. Parallel plate with overall and microscopic views for (a) smooth and (b) cross-hatched rough surfaces.**

The stress-ramp experiment on rough surfaces shows initial elastic deformation followed by a sudden increase in shear rate i.e. a stress plateau (Figures 8(a-d)). This is the classical solid-liquid transition stress referred to as static yield stress $\tau_y$ [48]. However, at a later segment the sudden rise in shear stress can be attributed to the fragmentation of the gel network [49]. For smooth surfaces, early yielding can be located. This is referred to as the critical shear stress for wall-slip $\tau_c$. However, the stress plateau for rough or smooth surfaces indicates the presence of shear banding in complex fluids [1]. Deducing data from Figures 8(a-d), we plotted $\tau$ versus the difference between the apparent shear rate $\dot{\gamma}_{app}$ (calculated for the cases of rough surfaces) and the nominal shear rate $\dot{\gamma}$ (calculated for the cases of smooth surfaces) to calculate slip-velocity $u_s$, and parameters $B$ and $m$ in Eq. (1) of the letter. Figures 8(a-d) suggest that $m$ varies from 1.5 to 3, depending upon the type of fluids. For a model waxy oil, $m$ varies from 2.2 to 2.8 with increasing wax concentration. The variable $B$ depends on kinetic parameters, and for isothermal study it is a constant with an order varying from $10^{-5}$ m Pa$^{-1}$s$^{-1}$ at $m = 1$ to $10^{-17}$ m Pa$^{-3}$s$^{-1}$ at $m = 3$. The variation of $m$ and $B$ is consistent with the literature [7-9, 37]. Figures 8(a-d) indicates that $\tau_c/\tau_y$ evolves in a manner that it satisfies the range of parameters assumed for the simulations in our letter. Furthermore, the flow curves at shear-rate controlled experiments in Figures 9(a, b) can be qualitatively tallied with the results for stress-ramp experiments for the determination of $\tau_y$ and $\tau_c$.

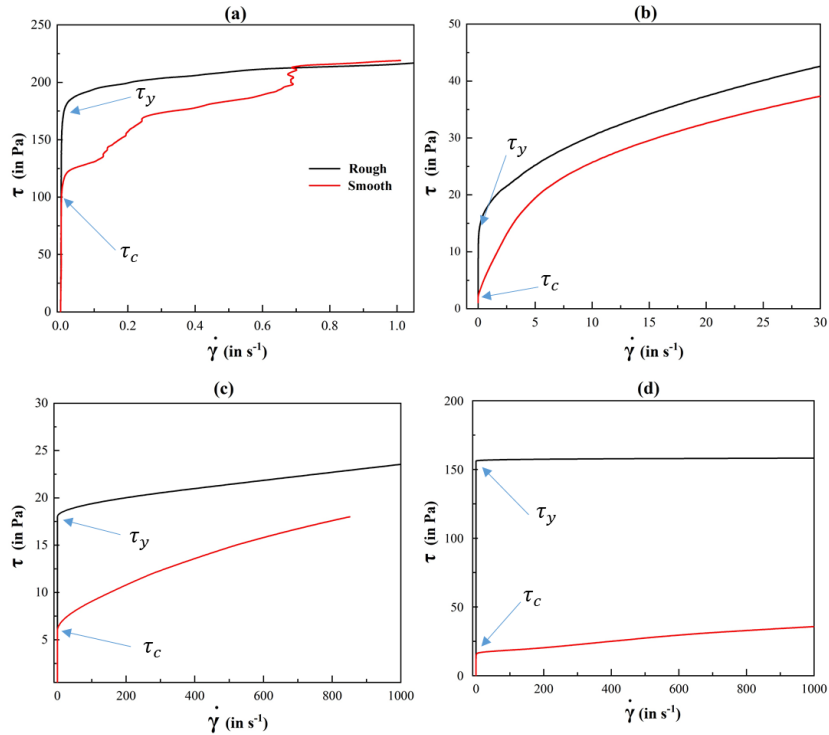

**Figure 8. Stress ramp experiments showing variations in stress $\tau$ with strain rate $\dot{\gamma}$ for different types of surfaces (rough and smooth) for (a) commercial toothpaste, (b) 1.5% Carbopol solution, (c) 6% model waxy oil, and (d) 7.5% model waxy oil**

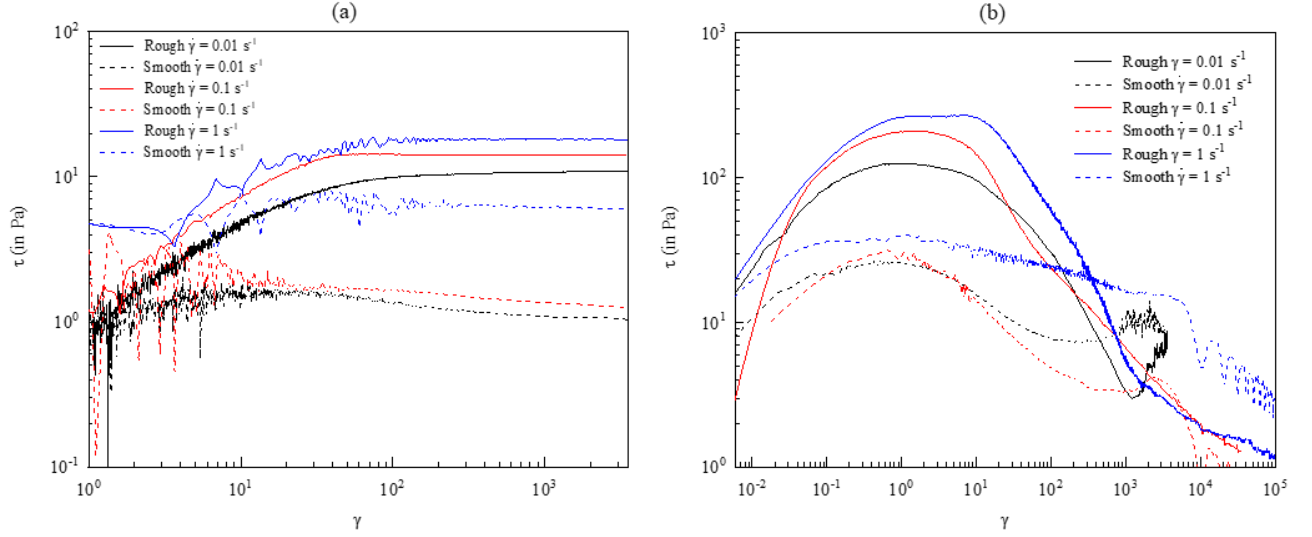

**Figure 9. Results for flow curve from shear-rate $\dot{\gamma}$ controlled experiments showing variation of stress $\tau$ with strain $\gamma$ for different types of surfaces (rough and smooth) for (a) 1.5% Carbopol solution and (b) 7.5% model waxy oil.**

**Governing equations, scaling and solution methodology**

The set of equations (1) to (5) are scaled to accommodate parameters like aspect ratio with less low simulation time while solving for series of non-linear partial differential equations. The axial coordinate and radial coordinates are scaled based on the length $L$ and radius $R$ of the pipeline as follows: $z' = z/L$, $r' = r/R$. The standardized velocity $W_s$ used for scaling of axial and radial velocities is calculated based on the magnitude of static yield stress $\tau_y$ as $W_s = R\tau_y/2\mu_0$. Therefore the axial and radial velocity components are written as $u' = u/\varepsilon W_s$ and $w' = w/W_s$, respectively. The critical length $L_c$ till which the flow always restart for a yield-stress fluid is defined as $L_c = PR/2\tau_y$. The dimensionless variable related to the aspect ratio of the pipeline $\varepsilon = R/L_c$ and a factor $\alpha$ which defined the ratio between the actual pipe length to the critical length, denoted as $\alpha = L/L_c$. The scaling of time $t'$ is done by resolving the smallest time scale phenomena, i.e., the compressive pressure wave propagation during initial stage of the flow restart [25, 50]. The scaled pressure and time are represented as $p' = p/P$ and $t' = t/(L_c\sqrt{\delta}/W_s)$ respectively. The viscosity is scaled based on the viscosity $\mu' = \frac{\mu}{\mu_\infty(P/\tau_y)}$. Finally, dimensionless numbers like modified Reynolds number $Re^*$ and compressibility number $\delta$ are used in the present study which helps in rewriting the governing equations in non-trivial form which can be used for larger parametric analysis. These dimensionless numbers are written as follows: $Re^* = \frac{\rho_0 R W_S}{\mu_\infty (P/\tau_y)}$, $\delta = P\kappa_\theta$. Finally, the shear or elastic modulus $G$ is scaled by a factor of $P/2$ and the dimensionless form of strain $\gamma'$ remains the same as the dimensional form.

The governing equations are written in dimensionless forms as follows:

Mass conservation equation:

$$\sqrt{\delta}\left(\frac{\partial p'}{\partial t'}\right) + \sqrt{\delta}\left[u'\frac{\partial p'}{\partial r'} + \frac{w'}{\alpha}\frac{\partial p'}{\partial z'}\right] + \frac{1}{r'}\frac{\partial(r'u')}{\partial r'} + \frac{1}{\alpha}\frac{\partial w'}{\partial z'} = 0 \dots\dots\dots\dots (1S).$$

Axial momentum conservation equation:

$$\frac{\partial w'}{\partial t'} + \sqrt{\delta}\left(u'\frac{\partial w'}{\partial r'} + \frac{w'}{\alpha}\frac{\partial w'}{\partial z'}\right) = -\frac{1}{\alpha\varepsilon Re^*}\frac{\partial p'}{\partial z'} + \frac{\varepsilon}{2Re^*}\left(\frac{1}{r'}\frac{\partial(r'\tau_{rz}')}{\partial r'} + \frac{1}{\alpha}\frac{\partial(\tau_{zz}')}{\partial z'}\right) \dots\dots\dots\dots (2S).$$

Radial momentum conservation equation:

$$\frac{\partial u'}{\partial t'} + \sqrt{\delta}\left(u'\frac{\partial u'}{\partial r'} + \frac{w'}{\alpha}\frac{\partial u'}{\partial z'}\right) = -\frac{1}{\varepsilon^2 Re^*}\frac{\partial p'}{\partial r'} + \frac{1}{2\varepsilon Re^*}\left(\frac{1}{r'}\frac{\partial(r'\tau_{rr}')}{\partial r'} + \frac{1}{\alpha}\frac{\partial(\tau_{rz}')}{\partial z'} - \frac{\tau_{\theta\theta}'}{r'}\right) \dots\dots\dots (3S).$$

Strain evolution equation for each component of strain tensor ($\gamma_{rz}', \gamma_{zz}', \gamma_{rr}', \gamma_{\theta\theta}'$):

$$\frac{1}{\sqrt{\delta}}\frac{\partial \gamma_{rz}'}{\partial t'} + \left(u'\frac{\partial \gamma_{rz}'}{\partial r'} + \frac{w'}{\alpha}\frac{\partial \gamma_{rz}'}{\partial z'}\right) - \left(\frac{\gamma_{rr}'}{\varepsilon}\frac{\partial w'}{\partial r'} + \frac{\gamma_{rz}'}{\alpha}\frac{\partial w'}{\partial z'} + \gamma_{rz}'\frac{\partial u'}{\partial r'} + \frac{\varepsilon\gamma_{zz}'}{\alpha}\frac{\partial u'}{\partial z'}\right) = \dot{\gamma}_{rz}' \dots\dots\dots (21),$$

$$\frac{1}{\sqrt{\delta}}\frac{\partial \gamma_{zz}'}{\partial t'} + \left(u'\frac{\partial \gamma_{zz}'}{\partial r'} + \frac{w'}{\alpha}\frac{\partial \gamma_{zz}'}{\partial z'}\right) - 2\left(\frac{\gamma_{rz}'}{\varepsilon}\frac{\partial w'}{\partial r'} + \frac{\gamma_{zz}'}{\alpha}\frac{\partial w'}{\partial z'}\right) = \dot{\gamma}_{zz}' \dots\dots\dots (4S),$$

$$\frac{1}{\sqrt{\delta}}\frac{\partial \gamma_{rr}'}{\partial t'} + \left(u'\frac{\partial \gamma_{rr}'}{\partial r'} + \frac{w'}{\alpha}\frac{\partial \gamma_{rr}'}{\partial z'}\right) - 2\left(\gamma_{rr}'\frac{\partial u'}{\partial r'} + \frac{\varepsilon\gamma_{rz}'}{\alpha}\frac{\partial u'}{\partial z'}\right) = \dot{\gamma}_{rr}' \dots\dots\dots (5S),$$

$$\frac{1}{\sqrt{\delta}}\frac{\partial \gamma_{\theta\theta}'}{\partial t'} + \left(u'\frac{\partial \gamma_{\theta\theta}'}{\partial r'} + \frac{w'}{\alpha}\frac{\partial \gamma_{\theta\theta}'}{\partial z'}\right) - 2\left(\gamma_{\theta\theta}\frac{u'}{r'}\right) = \dot{\gamma}_{\theta\theta}' \dots\dots\dots (6S).$$

where $\dot{\gamma}_{rz}', \dot{\gamma}_{zz}', \dot{\gamma}_{rr}', \dot{\gamma}_{\theta\theta}'$ are the components of strain rate tensor. Furthermore, the equation for dimensionless extra stress tensor is written as:

$$\ddot{\boldsymbol{\tau}}' = 2\mu'\ddot{\boldsymbol{d}}' - \frac{2}{3}\mu'(\boldsymbol{\nabla}' \cdot \boldsymbol{U}')\boldsymbol{I} + G'\ddot{\boldsymbol{\gamma}}' \dots\dots\dots\dots (7S).$$

**Boundary Conditions**

In the present problem, we consider gel degradation only after a pressured fluid is inserted at the inlet. The boundary conditions based on Dirichlet's and Neumann's convention are imposed for dimensionless pressure $p'$, strain, radial velocity $u'$, axial velocity $w'$ and extra shear stress $\tau'$ as follows:

At inlet: $p' = 1, u' = 0, \tau_{zz}' = 0$ and the pressurized fluid at the inlet has a strain $\gamma' = \gamma_{in}'$.

At outlet: $p' = 0, u' = 0, \tau_{zz}' = 0$ and constant flux condition is applied for strain; $\frac{\partial \gamma'}{\partial z'} = 0$.

Symmetric conditions prevail at the axis of the pipeline with no flow across the axis causing $u' = 0, \tau_{rz}' = 0$ and similar to the outlet boundary, Neumann's condition is applied for pressure and strain; $\frac{\partial p'}{\partial r'} = \frac{\partial \gamma'}{\partial r'} = 0$.

At the upper wall, slip based boundary conditions are set: $w' = u_s', u' = 0$. The wall-slip phenomenon in the present problem is isotropic in nature due to smooth walls [51].

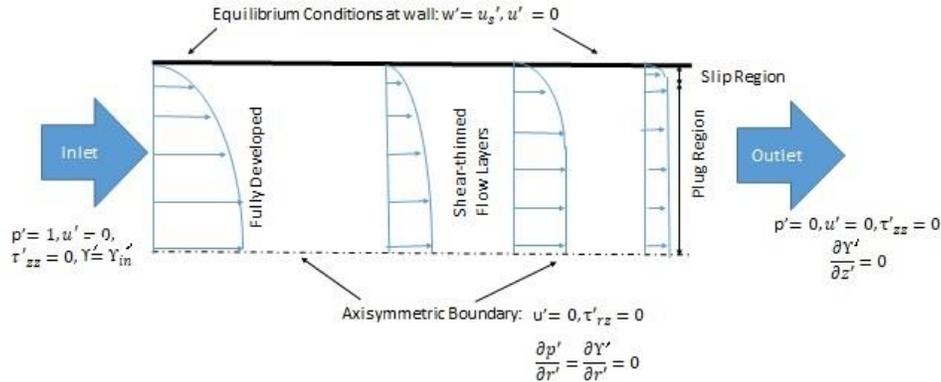

**Figure 10. Schematic diagram for flow representation**

**Solution methodology**

In the present problem, we considered a uniform and orthogonal staggered grid arrangement to represent the numerical domain $\Omega$. The governing equations are subjected to boundary conditions to solve for the primary variables $(u', w', p')$ and components of strain $(\gamma_{rz}', \gamma_{zz}', \gamma_{rr}', \gamma_{\theta\theta}')$ using finite volume methodology. Central difference scheme is applied for spatial discretization of velocity, pressure, stress, viscosity and strain-based components. The transient formulations are done using second-order implicit method. The staggered grid arrangement comprises of flux-related variables like velocity at the face centers of the cell; whereas, the properties like pressure is calculated at the volumetric center of the cell. The partial difference equations, thus formed after discretization, are solved using point-by-point iterative technique. The convergence criteria for velocity and pressure based variables are maintained at a dimensionless value of $10^{-12}$. A higher value for convergence criteria can lead to the divergence of the solution and a lower value of convergence criteria incurs high simulation time in addition to possibility of round-off errors. It is to be noted that the degradation rate constant $k$ in the present study is varied from 10 to 200. However, unless otherwise specified, the simulations are carried out primarily at $\delta = 4 \times 10^{-4}$, $k = 100$ and $\alpha = 1.5$.

## Data availability

The data sets generated during and/or analyzed during the current study are available from the corresponding author on reasonable request.

## Code availability

The codes of the computer simulations are available from the corresponding author upon request.

**Supplementary to the article "Synergetic Effect of Wall-Slip and Compressibility During Startup Flow of Complex Fluids" by A. Sanyal, S.B. Shinde, L. Kumar**

**1. Grid independence and model verification**

The benchmark for grid arrangement for the present problem is obtained from one of our previous studies [1]. In the present study, we compared several grid arrangements with number of grid points varying from 100×10 to 400×100 along axial ($N_z$) × radial ($N_r$) direction. The solutions at 200×20 is seen to be numerically adequate with the finest grid arrangement of 400×100. In the present problem, the wall-slip velocity near the wall is compared for numerical adequacy. Furthermore, the time-step of $10^{-4}$ following CFL criteria is taken as a benchmark for time-step independence tests. After comparisons of results for wall-slip velocity at two time instants ($t'$) for several values of time-step, $\Delta t' = 10^{-5}$ is finalized as numerically adequate.

The formulations for strain evolution based on upper-convection-derivative terms poses some intriguing issues based on the applicability of such formulation for a problem like the present one. Tikariha & Kumar [2] have shown that the upper-convection derivative based strain evolution methodology when compared to the ones adopted by [3] shows dissimilar results after the initial pressure propagation front has reached the outlet. However, in the present study, a similar verification is carried out at a longer pipe length. One can see that the results for pressure propagation at various instants of time remains same, irrespective of the type of formulation (Figure 1). The variations in results are marginal (< 1% relative deviation) in the regime of low compressibility number. Furthermore, the velocity profile (Figure 2) at a $\alpha z' = 0$, 0.75 and 1.5 for a combined effect of wall-slip and elasto-viscoplastic rheology shows variations from plug-like profile during gel degradation (at $t' < 100$ in Figures 2b and 2c) to a parabolic profile at a steady-state indicating Newtonian characteristics ($t' = 400$). This is qualitatively consistent with the numerical results for the velocity profiles of Damianou et al. [4]. In addition, the velocity magnitude at steady-state condition quantitatively complies with the analytical value obtained from Hagen–Poiseuille equation ($w_{max} = \frac{\Delta P}{4\mu L}R^2$). Finally, the code is subjected to the verification of the idea inspired from the experiments of El-Gendy et al. [5] which shows that the flow need not necessarily restart when the initial pressure propagation front reaches the outlet. One such scenario is shown in Figure 3 which occurs at $\alpha = 2, k = 50$ and $\tau_c = \frac{2}{3}\tau_y$ for an elasto-viscoplastic based rheology involving slip at $\delta = 4 \times 10^{-4}$.

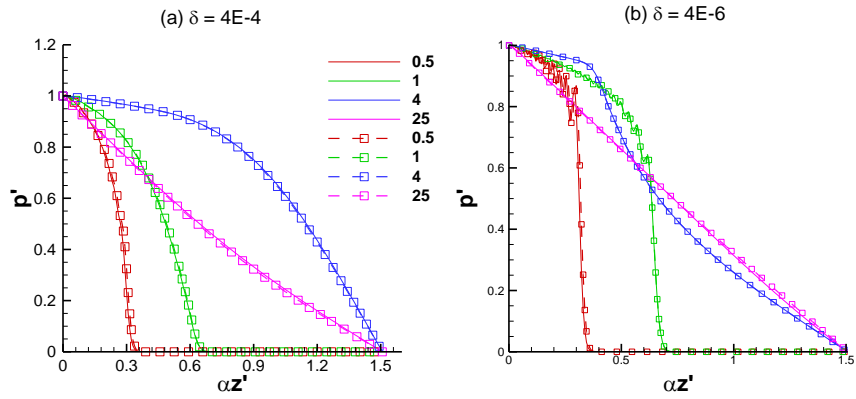

**Figure 1. Verification of pressure profiles from upper convection derivatives based strain evolution with strain evolution equation having only material derivative based strain evolution terms as used by Kumar et al. (2014) for different values of compressibility number (a) $\delta = 4 \times 10^{-4}$, and (b) $\delta = 4 \times 10^{-6}$ (at initial gel viscosity of 100 Pa s, *P* = 40 kPa and α = 1.5).**

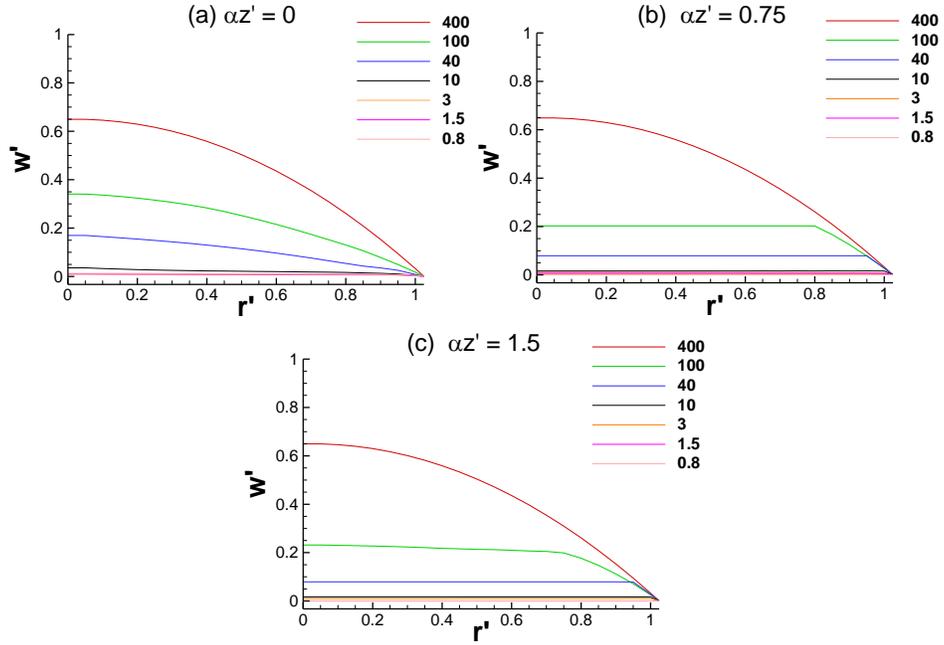

**Figure 2.** Time evolution of axial velocity variations along radial direction at different fixed axial locations (for initial gel viscosity of 100 Pa s, $P = 40$ kPa, $\delta = 4 \times 10^{-4}$, and α = 1.5) at $\tau_c=0$; (a) $\alpha z' = 0$, (b) $\alpha z' = 0.75$ and (c) $\alpha z' = 1.5$.

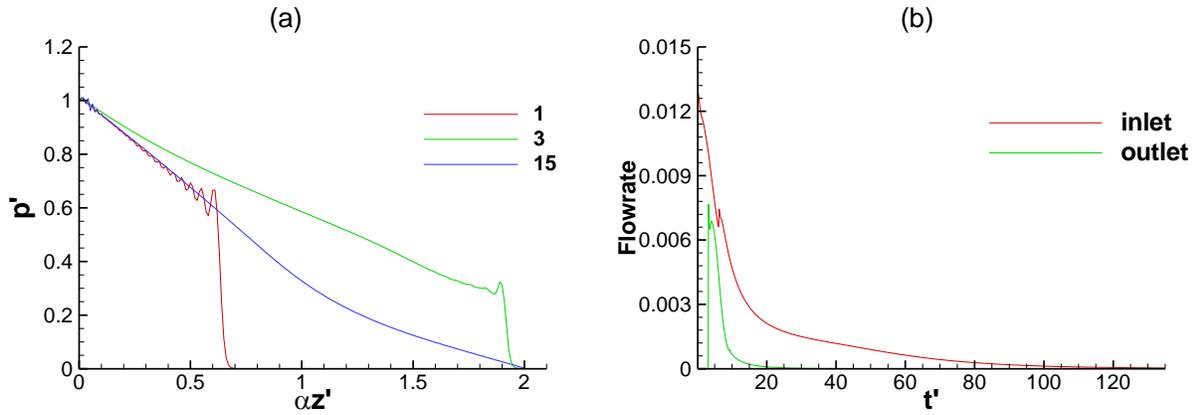

**Figure 3.** (a) Pressure propagation, and (b) time evolution of inlet and outlet flowrates; at initial gel viscosity of 100 Pa s, $\alpha = 2$, $P = 40$ kPa, $k = 50$, $\delta = 4 \times 10^{-4}$ and $\tau_c = 2/3\ \tau_y$.

**2. Time-dependent inlet and outlet flowrate variation at $\delta = 4 \times 10^{-4}$**

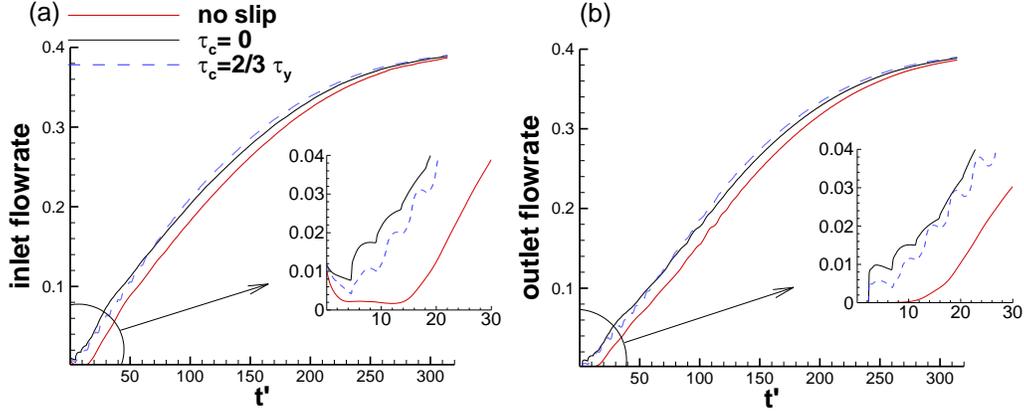

**Figure 4.** Comparison for (a) inlet and (b) outlet flowrates with time at initial gel viscosity of 100 Pa s, $\alpha = 1.5$, $P = 40$ kPa, $k = 100$, $\delta = 4 \times 10^{-4}$ with the cases for no-slip, shear-thinning slip ($\tau_c = 0$) and conditional slip scenario ($\tau_c = 2/3\ \tau_y$).

## 3. Pressure propagation at very low gel compressibility at $\delta = 4 \times 10^{-6}$

Figures 5(a-c) show pressure propagation from inlet to the outlet with marginal pressure drop along the axial direction in the downstream for a scenario involving wall-slip. For a wall-slip scenario, the overall gel deformation (combining net deformation at the bulk portion of the gel and the fluid-wall interface) during initial pressure propagation stage is less. This can be understood through negligible pressure gradient along the downstream and a sudden fall in pressure to zero at the outlet. Following a decay of transients, the no-slip scenario shows a linear profile at $t' = 40$ (Figure 5f).

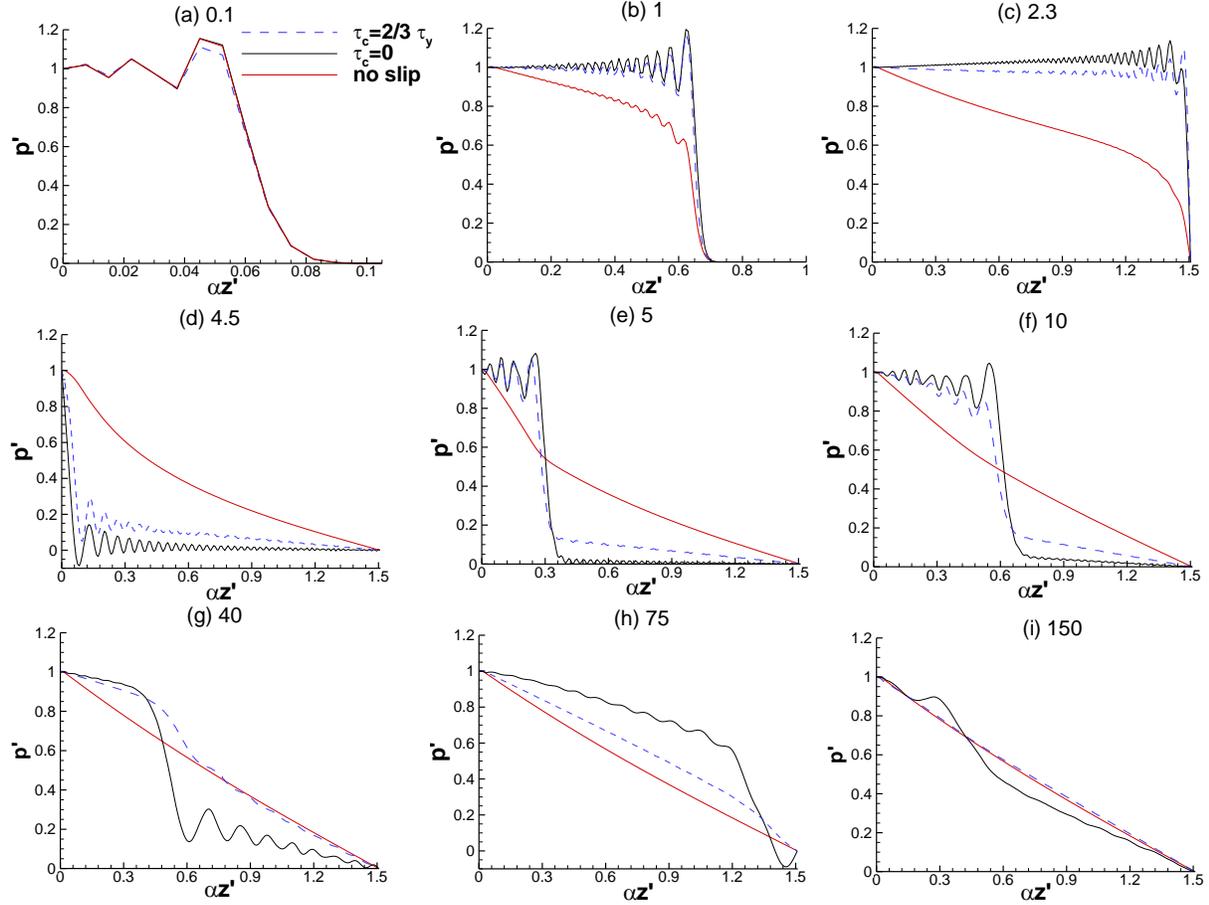

**Figure 5.** Comparison for the time evolution of axial pressure profile at initial gel viscosity of 100 Pa s, $\alpha = 1.5$, $P = 40$ kPa, $k = 100$, $\delta = 4 \times 10^{-6}$ with the cases for no-slip, shear-thinning slip ($\tau_c = 0$) and conditional slip scenario ($\tau_c = 2/3\ \tau_y$) at $t' =$ (a) 0.1, (b) 1, (c) 2.3, (d) 4.5 (e) 5, (f) 10, (g) 40, (h) 75 and (i) 150.

The reflected pressure wave causes a decrease in flow velocity compared to the flow induced from the approaching pressure front. It is to be noted that the reflected flow traverses back easily due to the slippage region without any significant viscosity attenuation at $\tau_c = 0$ (Figure 5d). At earlier time instants like $t' = 5$ (Figure 5e) for the cases at $\tau_c = 0$, the elongated trailing part in pressure propagation signal ($\alpha z' > 0.32$) denotes region of almost no gel deformation, and pressure signal moves back and forth via slip. At $\alpha z' < 0.3$, the inertial compressive pressure wave moves forward with an amplitude lower than the ones in Figure 5b. This indicates the gel movement with true velocity is causing viscous attenuation of pressure signal. A sudden drop in pressure signal at $0.3 < \alpha z' < 0.32$ specifies the region where the net resistive force from the gel's elastic strength is counterbalanced by the force associated with the approaching pressure wave (this region is analogous to a region of pressure front). However, for a subsequent span of time during the flow restart operation, the pressure wave continues to traverse through the slip region causing slower gel deformation and multiple reflection of pressure waves from the outlet to the inlet. The enhanced slope in the trailing part of the pressure profiles at subsequent time instants ($t' = 10$ (Figure 5f) and 40 (Figure 5g)) indicates gel degradation through viscous shearing forces from the interference between the backwardly reflected and forward approaching flow.

## 4. Wall-slip effects for pressure estimation during flow restart in longer pipelines and in scenarios for low gel degradation constants

In this case, we consider the overall aspect ratio ($1/\varepsilon$) of the pipeline as 400, which has a length equivalent to 4 times the critical length of the pipeline ($L_c$). Figure 6 shows that the initial pressure wave front reaches the outlet at $t' = 10$ at $\tau_c = 0$. Whereas, the pressure does not propagate beyond $\alpha z' = 3.4$ at $\tau_c = ^2/_3 \tau_y$ and $\alpha z' = 2.3$ at no-slip scenarios, respectively.

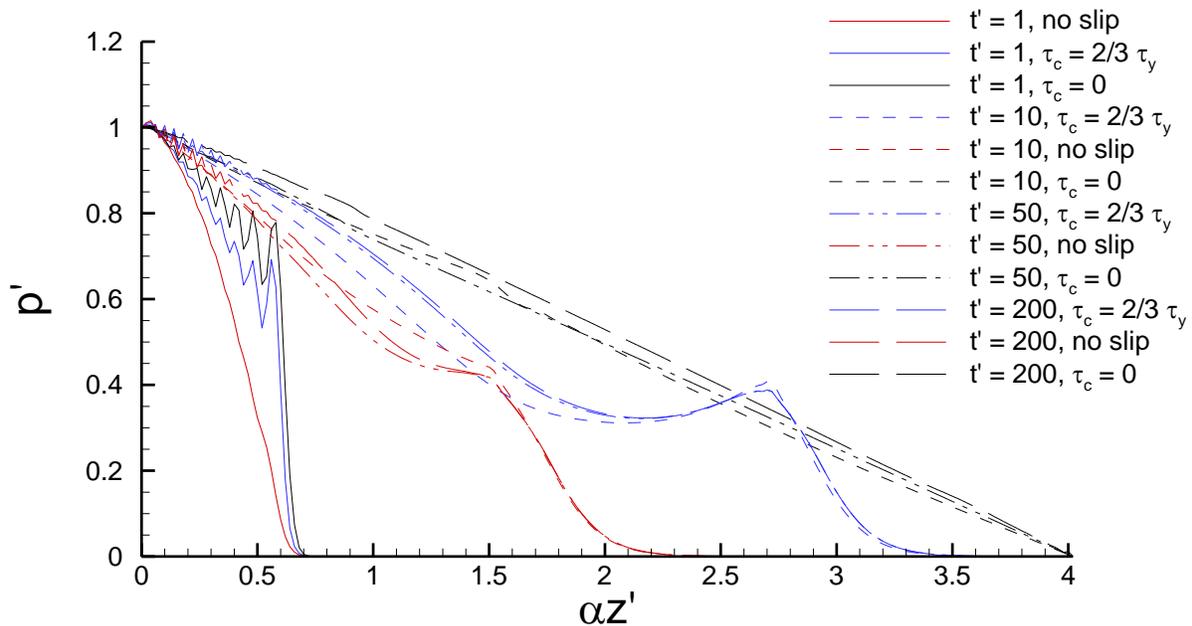

**Figure 6.** Comparison for time evolution of axial pressure propagation in longer pipeline for the cases of no-slip, $\tau_c = 0$, and $\tau_c = {}^2/_3\,\tau_y$ at the gel-wall interface for an initial gel viscosity 100 Pa s, $\alpha = 4$, $P = 40$ kPa and $\delta = 4 \times 10^{-4}$.

At low gel degradation rate, i.e. $k = 50$, the maximum yield strain required for beginning of the gel network's disengagement increases to 0.2 from 0.1 in the earlier-mentioned results. Unlike the case of no-slip (Figure 7a), the wall-slip allows pressure signals to propagate to the outlet for a pipeline having an overall aspect ratio ($1/\varepsilon$) of 200 or $\alpha = 2$ (Figures 11(b, c)). At $\tau_c = {}^2/_3\,\tau_y$, the initial pressure wave diffuses by compressing the gel network with minimum flow resistance near the wall. But, this does not guarantee a successful flow restart. One may see only inertial puncture at the outlet at $t' = 3$ in comparison to the fully-developed flow at the outlet at $t' = 10$. The non-linear pressure signals at late time instants like $t' = 500$ in Figure 7b suggests that the compressive resistances and viscous resistances will not be overcome in the long run, thereby leading to a halt in the flow. However, in Figure 7c, one may note the development of a linear profile at $t' = 50$, indicating the occurrence of complete gel compression and a tendency for steady-state flow at a later time.

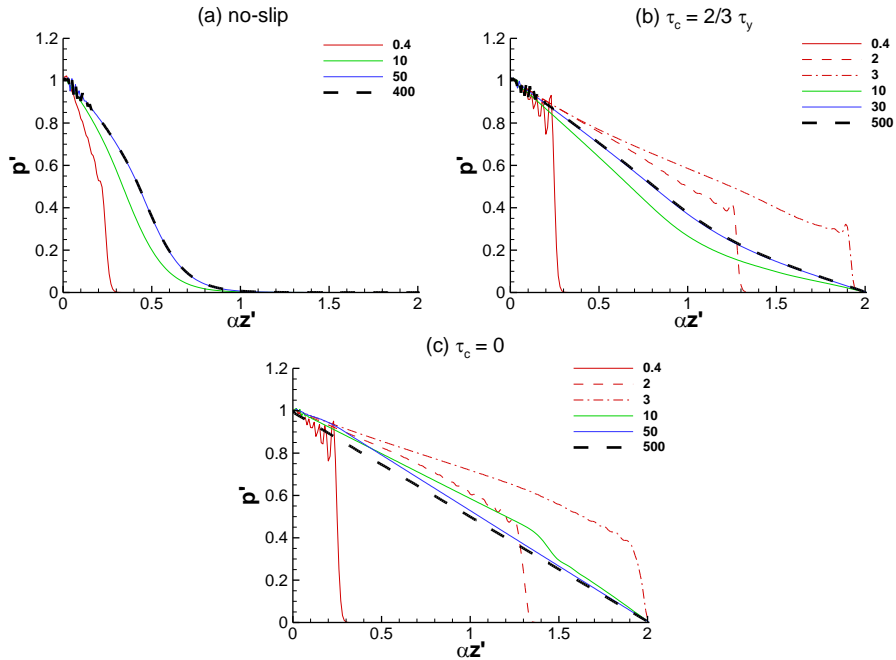

**Figure 7.** Comparison for time evolution of axial pressure propagation at low gel degradation scenario $k = 50$, for the cases of (a) no-slip, (b) $\tau_c = {}^2/_3 \tau_y$, and (c) $\tau_c = 0$ at the gel-wall interface; at initial gel viscosity 100 Pa s, $\alpha = 2$, $P = 40$ kPa and $\delta = 4 \times 10^{-4}$.

## 5. Local shear stress variations

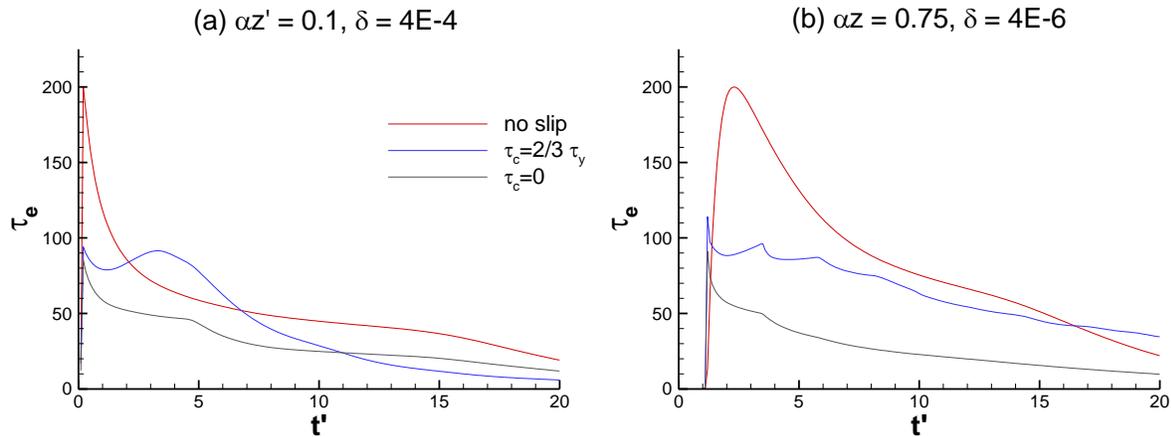

**Figure 8.** Comparison for the time evolution of true local stress $\tau_e$ between no-slip, conditional slip ($\tau_c = \frac{2}{3}\tau_y$) and shear-thinning slip ($\tau_c = 0$) at $r' = 0.975$ at the axial location (a) $\alpha z' = 0.1$ and (b) $\alpha z' = 0.75$; while the other gel conditions remaining the same as the ones mentioned in Figure 1 in the letter.